\documentclass[article]{elsarticle}

\usepackage{lineno,hyperref}
\usepackage[latin9]{inputenc}
\usepackage{amsmath}
\usepackage{amssymb}
\usepackage{graphicx}
\usepackage{subcaption}
\usepackage[section]{placeins}  

\newcommand{\sq}
{\nobreak \ifvmode \relax \else \ifdim\lastskip<1.5em
\hskip-\lastskip \hskip0.25em plus0em minus0.0em \fi \nobreak \vrule
height0.75em width0.75em depth0.0em\fi}
\usepackage{color}

\def\cC{\mbox{${\bf{C}}$}}
\def\cD{\mbox{${\bf{D}}$}}

\def\cS{\mbox{${\bf{S}}$}}

\begin{document}

\begin{frontmatter}

\title{Transformation design of in-plane elastic cylindrical cloaks, concentrators and lenses}

\author{Michele Brun\fnref{DIMCM}}
\author{S\'ebastien Guenneau \fnref{UMI}}
\fntext[DIMCM]{Dipartimento di Ingegneria Meccanica, Chimica e dei Materiali, Universit$\grave{\mbox{a}}$ di
Cagliari, Cagliari I-09123, Italy}
\fntext[UMI]{UMI 2004 Abraham de Moivre-CNRS, Imperial College, London SW7 2AZ, UK}



\date{\today}

\begin{abstract}
We analyse the elastic properties of a class of cylindrical cloaks
deduced from linear geometric transforms ${\bf x}\rightarrow {\bf x}'$ in the framework of
the Milton-Briane-Willis cloaking theory [New Journal of Physics 8, 248, 2006]. 
More precisely, we assume that the mapping between displacement fields
${\bf u}({\bf x}) \rightarrow {\bf u}'({\bf x}')$
is such that ${\bf u}'({\bf x}')={\bf A}^{-t}{\bf u}({\bf x})$, where
${\bf A}$ is either the transformation gradient $F_{ij}=\partial x'_i/\partial x_j$ or
the second order identity tensor ${\bf I}$.
The nature of the cloaks under review can be three-fold: some of
them are neutral for a source located a couple of wavelengths away;
other lead to either a mirage effect or a field confinement when the
source is located inside the concealment region or within their
coated region (some act as elastic concentrators squeezing the
wavelength of a pressure or shear polarized incident plane wave in
their core); a last category of cloaks is classified as an elastic
counterpart of electromagnetic perfect cylindrical lenses.
The former two categories require either rank-4 elastic tensor
and rank-2 density tensor and additional rank-3 and 2
positive definite tensors (${\bf A}={\bf F}$) or a rank 4 elasticity tensor and a
scalar density (${\bf A}={\bf I}$) with spatially varying positive values.
However, the latter example further requires that all rank-4, 3 and 2 tensors be
negative definite (${\bf A}={\bf F}$) or that 
the elasticity tensor be negative definite (and non fully symmetric) as well as a negative scalar density
 (${\bf A}={\bf I}$).
We provide some illustrative numerical examples with the Finite
Element package Comsol Multiphysics when ${\bf A}$ is the identity.
\end{abstract}




\end{frontmatter}



\section{Introduction}
There has been a growing interest over the past years in the analysis of elastic waves in thin plates in the metamaterial community with the theoretical proposal \cite{farhat09,farhat09PRL},
and its subsequent experimental validation \cite{stenger12,misseroni16} of a broadband cloak for flexural waves. Square \cite{colquitt14} and diamond \cite{pomot2019form} cloaks are based on an improved transformed plate model, while form-invariance of the transformed equations in the framework of pre-stressed anisotropic plates is analized in \cite{Brun14b,morvaridi18,Golgoon21}.

There is currently a keen activity in transformation optics, whereby
transformation based solutions to the Maxwell equations expressed in
curvilinear coordinate systems travel along geodesics rather than in
straight lines \cite{rayoptics}. The fact that light follows
shortest trajectories, the physical principle behind transformation
optics, was formulated by de Fermat back in 1662. This minimization
principle is applicable to ray optics, when the wavelength is much
smaller than the size of the diffraction object. Leonhardt has shown
in 2006 \cite{leonhardt06} that this allows for instance the design
of invisibility cloaks using conformal mappings. Pendry, Schurig and
Smith simultaneously reported that the same principle applies to
electromagnetic waves, i.e. when the wavelength is in resonance with
the scattering object, by creating a hole in the curved space
\cite{pendry2006controlling}. Interestingly, the mathematicians Greenleaf,
Lassas and Uhlmann proposed an earlier route to invisibility using
an inverse problem approach in 2003 \cite{greenleaf03}, and together
with Kurylev have been able since then to bridge the cloaking theory
with Einstein theory of relativity, thereby suggesting possible
avenues towards electromagnetic wormholes \cite{greenleafprl07,kadic2014invisible}.
Leonhardt and Philbin have further proposed an optical fibre
experiment \cite{philbin08} for an analogue of Hawking's famous
event horizon in his theory of black holes \cite{hawking}. It seems
therefore fair to say that transformation optics offers a unique
laboratory for thought experiments, leading to a plethora of
electromagnetic paradigms. However, this would remain some academic
curiosity without the practical side effect: so-called
metamaterials, first introduced by Pendry in 1999 to obtain
artificial magnetism in locally resonant periodic
structures\cite{pendry1999magnetism}.

The first realization of an electromagnetic invisibility cloak
\cite{schurig06} is a metamaterial consisting of concentric arrays
of split-ring resonators. This structured material effectively maps
a concealment region into a surrounding shell thanks to its strongly
anisotropic effective permittivity and permeability which further
fulfil some impedance matching with the surrounding vacuum. The
cloak thus neither scatter waves nor induces a shadow in the
transmitted field. Split ring resonators enable to meet among others
the prerequisite artificial magnetism property, otherwise
unobtainable with materials at hand \cite{pendry1999magnetism}.
This locally
resonant micro-structured cloak was shown to conceal a copper
cylinder around $8.5$ GHz, as predicted by numerical simulations
\cite{schurig06}.

The effectiveness of the transformation based invisibility cloak was
demonstrated theoretically by Leonhardt \cite{leonhardt06} solving
the Schr\"odinger equation. Note that this equation is not only valid
to compute ray trajectories (geodesics) in the geometrical optic
limit, but also for matter waves in the quantum theory framework.
Zhang et al. used this analogy to propose a quantum cloak based upon
ultracold atoms within an optical lattice \cite{zhangprl08}.
Greenleaf et al. subsequently discussed resonances (so-called
trapped modes) occurring at a countable set of discrete frequencies
inside the quantum cloak, using a spectral theory approach
\cite{greenleafprl08}.

Using analogies between the Helmholtz and the Schr\"odinger
equations, Cummer and Schurig demonstrated that pressure acoustic
waves propagating in a fluid also undergo the same geometric
transform in 2D \cite{cummer06b}. Chen and Chan further extended
this model to 3D acoustic cloaks \cite{chen07}, followed by an
independent derivation of the acoustic cloak parameters in
\cite{cummer08}. Such meta-fluids require an effective anisotropic
mass density as in the model of Torrent and Sanchez-Dehesa
\cite{sanchez}. However, an acoustic cloak for linear surface water
waves studied experimentally and theoretically in \cite{farhat08},
only involves an effective anisotropic shear viscosity.

Nevertheless, transformation based invisibility cloaks cannot be
applied in general to elastodynamic waves in structural mechanics as
there is a lack of one-to-one correspondence between the equations
of elasticity and the Schr\"odinger equation \cite{milton06b}.
Bigoni et al. actually studied such neutral inclusions in the
elastostatic context using asymptotic and computational methods in
the case of anti-plane shear and in-plane coupled pressure and shear
polarizations \cite{bigoni98}, but when one moves to the area of
elastodynamics, geometrical transforms become less tractable and
neutrality breaks down: there are no conformal maps available in
that case, and one has to solve inherently coupled tensor equations.

More precisely, Milton, Briane and Willis have actually shown that there is no
symmetric rank 4 elasticity tensor describing the heterogeneous
anisotropic medium required for an elastodynamic cloak in the context of Cauchy elasticity
\cite{milton06b}. However, so-called Willis's equations, discovered by the British applied mathematician
John Willis in the early 80's \cite{willis1981variational,willis1985nonlocal}, offer a new paradigm for elastodynamic cloaking, as
they allow for introduction of additional rank-3 and rank-2 tensors in the equations of motion that
make cloaking possible. 

Nevertheless, Brun, Guenneau and Movchan have shown
\cite{brunapl} that it is possible to design an elastic cloak
without invoking Willis's equations for
in-plane coupled shear and pressure waves with a metamaterial
described by a rank 4 elasticity tensor preserving the main
symmetries, as well as a scalar density. Importantly, both
elasticity tensor and density are spatially varying, and the former
one becomes singular at the inner boundary of the cloak
\cite{brunapl}. Some design based on a homogenization approach for
polar lattices has been proposed by Nassar, Chen and Huang
\cite{nassar2018degenerate} and Garau et al. \cite{Garau2019}. Achaoui et al. have
proposed an alternative design making use of elastic swiss-rolls
\cite{achaoui2020cloaking}. Diatta and Guenneau \cite{diatta2014controlling} have shown that
a spherical elastodynamic cloak can be designed
using the same route as in \cite{brunapl},
but the corresponding metamaterial design remains an open problem.
There is an alternative, pre-stress, route to elastic cloaking
proposed by Norris and Parnell that greatly relaxes constraints on material properties compared
to the previous routes \cite{parnell2012nonlinear,norris2012hyperelastic,parnell2012employing}.

In the present article, we further investigate cylindrical cloaks
for in-plane elastic waves using a radially symmetric linear
geometric transform which depends upon a parameter. Depending upon
the value of the parameter, the transform is applied to the design
of neutral (invisibility) cloaks, elastic concentrators or
cylindrical lenses. We discuss their underlying mechanism using a
finite element approach which is adequate to solve the Navier
equations in anisotropic heterogeneous media.

\section{Governing equations and elastic properties of cloaks}
\subsection{The equations of motion}
The propagation of in-plane elastic waves is governed by the Navier equations. 
Assuming time harmonic $\exp(-i\omega t)$ dependence, with $\omega$ as the wave frequency, allows us to work directly in the spectral domain.
Such dependence is assumed henceforth and suppressed, leading to
\begin{equation}
\nabla\cdot{\bf C}:\nabla{\bf u}+\rho\,\omega^2 {\bf u}+{\bf b}={\bf 0} \; ,
\label{navier}
\end{equation}
where, considering cylindrical coordinates $(r,\theta)$, ${\bf u}=(u_r,u_\theta)$ is the in-plane displacement, 
$\rho$ the density and $C_{ijkl}$ $(i,j,k,l=r,\theta)$ the fourth-order elasticity tensor of the (possibly heterogeneous anisotropic) elastic medium.  
In eqn. (\ref{navier}) ${\bf b}$ is the body force.

\subsection{The transformed equations of motion}

\begin{figure}
\centering
\includegraphics[width=10cm]{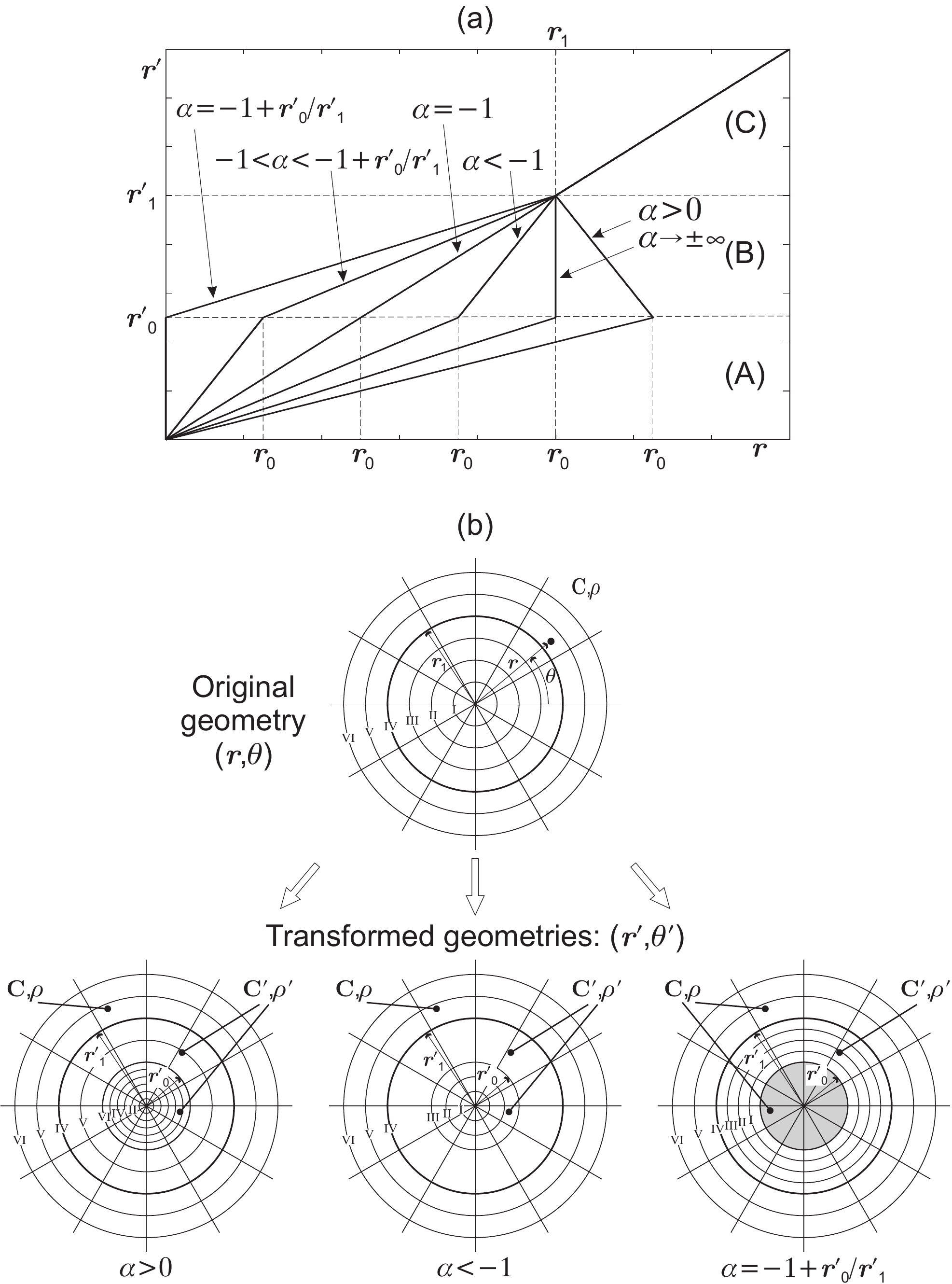}
\vspace{0.2cm}
\caption{Geometric transform of eqn. (\ref{PTransform}). (a) Representation of the transform $r\rightarrow r'$
for different values of the parameter $\alpha$. The domains A $(r'\leq r'_0)$, B $(r'_0\leq r'\leq r'_1)$ and
C $(r'\geq r'_1)$ are indicated. (b) Transformation of the geometry for $\alpha>0$, $\alpha<1$ and $\alpha=-1+r'_0/r'_1$ (perfect cloak).}
\label{fig01}
\end{figure}

Let us now consider the radial linear geometric transform $(r,\theta)\rightarrow(r',\theta')$, with
$\theta'=\theta$, shown in Fig. \ref{fig01}
\begin{equation}
r'\!=\!\left\{
\begin{array}{lcrl}
[(1\!+\!\alpha)r_1\!-\!\alpha r_0] \frac{r}{r_0} & \mbox{for} & r'\leq r'_0 & \mbox{(domain A)},\\
(1\!+\!\alpha)r_1\!-\!\alpha r & \mbox{for} & r'_0\leq\! r'\!\leq r'_1 & \mbox{(domain B)},\\
r & \mbox{for} & r'\geq r'_1 & \mbox{(domain C)},
\end{array}
\right.
\label{PTransform}
\end{equation}
where $\alpha=-(r'_1-r'_0)/(r_1-r_0)$ is a real parameter and $r'_0=(1+\alpha)r_1-\alpha\, r_0$, $r'_1=r_1$.
The transformation gradient is ${\bf F}=(dr'/dr){\bf I}_r+(r'/r){\bf I}_\bot$ where ${\bf I}_r={\bf e}_r\otimes{\bf e}_r$ 
is the second-order projection tensor along the radial direction identified by the unit vector ${\bf e}_r$, and ${\bf I}_\bot={\bf I}-{\bf I}_r$, with ${\bf I}$ second-order identity tensor.
Furthermore, $J=\det {\bf F}$ is the Jacobian of the transformation.
 
Design of in-plane transformation-based elastic cloaks has been
first discussed in \cite{brunapl} when $\alpha=-1+r'_0/r'_1$ ($r_0=0$): in that
case, (\ref{PTransform}) simplifies into the geometric transform for
an invisibility cloak $ r'=r'_0+ \frac{r'_1-r'_0}{r'_1} r$ in the domain (B)
\cite{pendry2006controlling,greenleaf03}, where $r'_0$ and $r'_1$, respectively,
denote the inner and outer radii of the circular cloak. However,
other values of the parameter $\alpha$ lead to equally interesting
cloaks, such as neutral concentrators, first studied in the context
of electromagnetism \cite{rahm}, and we would like to discuss these
in the sequel.

We now need to consider two cases for the transformed equations of motion.
 
\subsubsection{Gauge transform ${\bf u}'(r',\theta')={\bf u}(r,\theta)$}
By application of transformation (\ref{PTransform}) with the Gauge ${\bf u}'(r',\theta')={\bf u}(r,\theta)$
the Navier eqns.(\ref{navier}) are mapped into the equations
\begin{equation}
\nabla'\cdot{\cC'}:\nabla'{\bf u}'+\rho'\omega^2 {\bf u}'+{\bf b}'={\bf 0} \;,\label{snavier}
\end{equation}
where ${\bf u}'(r',\theta')$ and ${\bf b}'(r',\theta')$ are the transformed displacement and body force, respectively, 
and $\nabla'={\bf F}^t\nabla$ the gradient operator in the transformed coordinates. 
In particular, we stress that we assume an identity gauge transformation\cite{brunapl,NorrisShuvalov2011}, i.e. ${\bf u}'(r',\theta')={\bf u}(r,\theta)$. 
The stretched density is the scalar field
\begin{equation}
\rho'=
\left\{
\begin{array}{ll}
\displaystyle{\left[\frac{(1\!+\!\alpha)r'_1-r'_0}{\alpha r'_0}\right]^2 \rho} & \mbox{in A},\\[3.5 mm]
\displaystyle{\frac{r'-(1\!+\!\alpha)r'_1}{\alpha^2 r'} \rho} & \mbox{in B},\\[4.5 mm]
\rho & \mbox{in C},
\end{array}
\right.
\label{srho}
\end{equation}
homogeneous in A and C.
The transformed linear elasticity tensor has components 
\begin{equation}
\label{sc0}
C'_{ijkl}=J^{-1}C_{mnop}F_{im}F_{ko}\delta_{jn}\delta_{lp}\;,
\end{equation}
where $(i,j,k,l=r',\theta')$, $(m,n,o,p=r,\theta)$, $\delta_{jn}$ is the Kronecker delta and the usual summation convention over repeated indices is used. 
In particular, if before transformation the material is isotropic, i.e.
 $C_{ijkl}=\lambda\delta_{ij}\delta_{kl}+\mu(\delta_{ik}\delta_{jl}+\delta_{il}\delta_{jk})$ $(i,j,k,l=r,\theta)$, with $\lambda$ and $\mu$ the Lam\'e moduli, the transformed elasticity tensor $\cC'$ has non zero cylindrical components
\begin{equation}
\begin{array}{ll}
C'_{r'\!r'\!r'\!r'\!}\!=\!\frac{r'-(1+\alpha)r_1}{r'}(\lambda\!+\!2\mu), \!&
C'_{\!\theta'\!\theta'\!\theta'\!\theta'}\!=\!\frac{r'}{r'-(1+\alpha)r_1}(\lambda\!+\!2\mu),\\[1.6mm]
C'_{r'\!r'\!\theta'\!\theta'\!}=C'_{\theta'\!\theta'\! r'\!r'}=\lambda, &
C'_{r'\!\theta'\!\theta' \!r'}=C'_{\theta'\! r'\!r'\!\theta'}=\mu, \\[1.6mm]
C'_{r'\!\theta'\! r'\!\theta'}=\frac{r'-(1+\alpha)r_1}{r'} \mu, &
C'_{\theta'\! r'\!\theta'\! r'}=\frac{r'}{r'-(1+\alpha)r_1} \mu \; ,
\end{array}
\label{sc}
\end{equation}
in B and $\cC'=\cC$ in A and C.

The transformation and the corresponding transformed density $\rho'$ and elasticity tensor $\cC'$ are broadband, they do not depend 
on the applied frequency $\omega$.

\subsubsection{Gauge transform ${\bf u}'(r',\theta')={\bf F}^{-t}{\bf u}(r,\theta)$}
As noted in \cite{milton06b}, by application of transformation (\ref{PTransform}) with the Gauge ${\bf u}'(r',\theta')={\bf F}^{-t}{\bf u}(r,\theta)$, where ${\bf F}$ is the transformation gradient,
the Navier eqns.(\ref{navier}) are mapped into the equations
\begin{equation}
\begin{array}{ll}
\label{eq:willis-model-change-of-variable-princ}
\nabla' \cdot \Big(\cC'': \nabla' {\mathbf u}' + \cD' \cdot {\mathbf u}' \Big) + \cS' : \nabla' {\mathbf u}' + \omega^2 \rho' \mathbf{u}' + \mathbf{b}' = \mathbf{0} \; .
\end{array}
\end{equation}

The transformed rank-4 elasticity tensor $\cC''$ has components 
\begin{equation}
\label{scwill}
C''_{ijkl}=J^{-1}F_{im}F_{jn}C_{mnop}F_{ko}F_{lp} \; ,
\end{equation}
where $(i,j,k,l=r',\theta')$, $(m,n,o,p=r,\theta)$.

We note that $\cC''$ in (\ref{scwill}) has all the symmetries, unlike $\cC'$ in (\ref{sc}), which has the major but not the minor simmetries.

The rank-3 tensors $\cD'$ and $\cS'$ in (\ref{scwill}) have elements
\begin{equation}
\label{sDwill}
D'_{ijk}=J^{-1}F_{im}F_{jn}
C_{mnop}
\frac{\partial^2 x'_k}{\partial x_o \partial x_p}
=
D'_{jik}
 \; ,
\end{equation} 
and
\begin{equation}
\label{sSwill}
S'_{ijk}=J^{-1}
\frac{\partial^2 x'_i}{\partial x_m \partial x_n}
C_{mnop}
F_{jo}F_{kp}
=S'_{jik}
\; .
\end{equation}

Finally, the transformed density $\rho'$ in (\ref{scwill}) is matrix valued
\begin{equation}
\label{srhowill}
\rho'_{ij}=J^{-1}\rho F_{im}F_{jm}+J^{-1} \frac{\partial^2 x'_i}{\partial x_m \partial x_n}
C_{mnop}
\frac{\partial^2 x'_j}{\partial x_o \partial x_p}
=
\rho'_{ji}
 \; .
\end{equation} 
These expressions were first derived in \cite{milton06b}.

Now, if before transformation the material is isotropic,
then the transformed elasticity tensor $\cC''$ has non zero cylindrical components
\begin{equation}
\begin{array}{l}
C''_{r'\!r'\!r'\!r'\!}\!=\!\frac{r'-(1+\alpha)r_1}{r'(r_1-r_0)^2}(\lambda\!+\!2\mu),
\\[1.6mm]
C''_{\!\theta'\!\theta'\!\theta'\!\theta'}\!=\!\frac{r'^3}{(r_1-r_0)^2(r'-(1+\alpha)r_1)^3}(\lambda\!+\!2\mu),
\\[1.6mm]
C''_{r'\!r'\!\theta'\!\theta'\!}=C''_{\theta'\!\theta'\! r'\!r'}=\frac{r'}{(r_1-r_0)^2(r'-(1+\alpha)r_1)}\lambda,
\\[1.6mm]
C''_{r'\!\theta'\!\theta' \!r'}=C'_{\theta'\! r'\!r'\!\theta'}
=C''_{r'\!\theta'\! r'\!\theta'}=C''_{\theta'\! r'\!\theta'\! r'}=
\frac{r'}{(r_1-r_0)^2(r'-(1+\alpha)r_1)} \mu \; ,
\end{array}
\label{scwillwill}
\end{equation}
in B and $\cC''=\cC$ in A and C.

On the other hand, the rank-3 tensors $\cD'$ and $\cS'$ have non zero cylindrical components
\begin{equation}
\begin{array}{l}
D'_{r'\!r'\!r'\!}\!=\frac{1}{(r_1-r_0)^2(r'-(1+\alpha)r_1)(r'+r_1r_0/(r_1-r_0))}\lambda
=-S'_{r'\!r'\!r'}\!
\\[1.6mm]
D'_{r'\!\theta'\!\theta'}=D'_{\theta'\! r'\!\theta'}
=2\frac{1}{(r_1-r_0)^2(r'-(1+\alpha)r_1)^2} \mu
=-S'_{\theta'\!\theta'\! r'}=-S'_{\theta'\! r'\!\theta'}
\\[1.6mm]
D'_{\theta'\!\theta'\! r'}\!=\frac{r'+r_1r_0/(r_1-r_0)}{(r_1-r_0)^2(r'-(1+\alpha)r_1)^3}(2\mu+\lambda)
=-S'_{r'\!\theta'\!\theta'}\!
\; .
\end{array}
\label{sdwillwill}
\end{equation}

Similar expressions can be derived for the transformed density. Expressions in (\ref{scwillwill}) and \ref{sdwillwill}) are more intricate than those in (\ref{sc}); thus, in the sequel, we focus on the transformed equations of motion (\ref{snavier}). 

\subsection{Interface conditions for Gauge ${\bf u}'(r',\theta')={\bf u}(r,\theta)$}
\label{SectionIC}

Perfect cloaking and perfect concentrator require additional conditions on displacements and tractions at the interfaces between the domains with different material properties introduced by the transformation (\ref{PTransform}).
In the transformed problem (\ref{snavier}) there are two interfaces, between domains A and B, at $r'=r'_0$ and $r=r_0$, and 
at the cloak's outer boundary, between domains B and C, at $r'=r=r_1$.
Transformed equations (\ref{snavier}) together with the assumption ${\bf u}'(r',\theta')={\bf u}(r,\theta)$ assure that displacements and tractions in the inhomogeneous transformed domain at a point $(r',\theta')$ coincide with displacements and tractions at the corresponding point $(r,\theta)$ in the original homogeneous problem (\ref{navier}) where no interfaces between different materials are  present.
In particular, if we introduce the Cauchy stress tensors ${\bf \sigma}'=\cC'\!:\!\nabla'{\bf u}'$ and ${\bf \sigma}=\cC\!:\!\nabla{\bf u}$ for the transformed and original problem, respectively,
it is verified the following equality between tractions
\begin{equation}
{\bf \sigma}'\cdot{\bf e}_r'={\bf \sigma}\cdot{\bf e}_r,
\label{EqTrac}
\end{equation}
at $r'=r'_0$, $r=r_0$ and at $r'=r=r_1$.
Equality (\ref{EqTrac}) can be easily demonstrated by using Nanson's formula
\cite{Ogden1997} ${\bf e}_r'=J {\bf F}^{-t}{\bf e}_r$ for the radial unit vectors.

Note that the matching is independent on the particular value assumed by $\alpha$.

\subsection{Perfect cloak. Singularity at the inner interface}

We note that for the perfect cloak\cite{brunapl}, i.e. $r_0=0$ and $\alpha=-1+r'_0/r'_1$, a point at $r=0$ is mapped into a disk of radius $r'_0$.
This is a singular transformation and, at the cloak inner boundary, $r'-(1+\alpha)r_1\to 0$.
Therefore, at  $r'=r'_0$, from (\ref{srho}) and (\ref{sc}) one can see that $\rho'\to 0$, $C'_{r'\!r'\!r'\!r'\!}\,,\, C'_{r'\!\theta'\! r'\!\theta'}\to 0$, while $C'_{\theta'\!\theta'\! \theta'\!\theta'}\,, C'_{\theta'\! r'\!\theta'\!r'}\to \infty$.

Similarly,  at  $r'=r'_0$, from (\ref{scwillwill}) and (\ref{sdwillwill}) one can see that $C''_{r'\!r'\!r'\!r'\!}\to 0$, while $C''_{\theta'\!\theta'\! \theta'\!\theta'}\,,C''_{r'\!r'\!\theta'\!\theta'\!}=C''_{\theta'\!\theta'\! r'\!r'} \,, C''_{r'\!\theta'\!\theta' \!r'}=C'_{\theta'\! r'\!r'\!\theta'}
=C''_{r'\!\theta'\! r'\!\theta'}=C''_{\theta'\! r'\!\theta'\! r'}\to \infty$. Moreover, one notes that the rate of divergence is faster for ${\bf C''}$ than ${\bf C}'$, and thus anisotropy is even more extreme in the neighborhood of the inner boundary for ${\bf C''}$.
All expressions in (\ref{sdwillwill}) diverge when $r'-(1+\alpha)r_1\to 0$.    

The required extreme anisotropy physically means that pressure and
shear waves propagate with an infinite velocity in the azimuthal
$\theta'$-direction and zero velocity in the radial $r'$-direction along the inner boundary, which results in a
vanishing phase shift between a wave propagating in a homogeneous
elastic space and another one propagating around the coated region.

Clearly the presence of unbounded physical properties poses limitations on possible realizations and numerical implementation of the model; regularization techniques have been proposed introducing the concept of {\em near cloak} \cite{Kohn2008,Colquittetal2013,Jonesetal2015},
but the realization of such elastodynamic cloaks remains a challenge.

\subsection{General transformation}

We now wish to extend first the proposal of Rahm et al. \cite{rahm} of an omni-directional electromagnetic concentrator to the
elastic setting and second to consider a more general transformation including folded transformed geometries, as proposed by 
for quasi static equations of electromagnetism by Milton et al. \cite{Miltonetal2008}. 
We recall that the transformation (\ref{PTransform}) compress/expand a disc with radius $r_0$ at the expense of an expansion/compression of the annulus between $r_0$ and $r_1$.
The inner disk is expanded for $-1<\alpha<-1+r_0'/r_1'$ with the limiting cases $\alpha=-1$ corresponding to an identity and $\alpha=-1+r_0'/r_1'$ to perfect cloaking. On the contrary
the disk is compressed for $\alpha<-1$ and $\alpha>0$, and, in the case of positive value of the topological parameter $\alpha$, $r_0>r_1$ and a folding of the original geometry is obtained.

The material remains homogeneous and isotropic in the inner disk A where only the density is changed. In the annulus region B the material is heterogeneous and elastically anisotropic. Consistently with Brun et al.\cite{brunapl} and differently from Milton et al. \cite{milton06b} the density remains a scalar field.
We stress that the heterogeneity is smoothly distributed and the material is functionally graded with the absence of any jump in the material properties leading to possible scattering effects.
As detailed above interface conditions are automatically satisfied and do not introduce any scattering.

It is important to note that, excluding the perfect cloaking case, for bounded values of $\alpha$ all the elastic rigidities and the density are bounded leading to possible physical and numerical implementation of the metamaterial structure.

\subsection{The radial field concentrator. Unbounded $\alpha$}

The limiting cases $\alpha\to \pm\infty$, where $r_0=r_1$, correspond asymptotically to the radial transformation
\begin{equation}
r'=
\left\{
\begin{array}{ll}
\frac{r}{r_1} & \mbox{in A},\\
r_1 & \mbox{in B},\\
r & \mbox{in C}.
\end{array}
\right.
\label{eqn1001}
\end{equation}

In such a case in the annulus region B $C'_{\theta'\!\theta'\! \theta'\!\theta'}\,, C'_{\theta'\! r'\!\theta'\!r'}\to \infty$, $\rho'\to\infty$ and all the other elastic rigidities components are unchanged. In such material, independently on the external field impinging the metamaterial region $r'\le r'_1$, the elastic fields in B are radially independent and depend only on the azimuthal coordinate $\theta'$. However, the harmonic behavior cannot be reached in a finite time after the transient since the density $\rho'$ is unbounded.

\section{Numerical results and discussion}

In this section, we report the finite element computations performed
in the COMSOL multiphysics package. Normalized material parameters are used. A cloak of density $\rho'$
(\ref{srho}) and elasticity tensor $\cC'$ (\ref{sc0},\ref{sc}) is embedded in
an infinite isotropic elastic material with normalized Lam\'e moduli
$\lambda=2.3$ and $\mu=1$, and mass density $\rho=1$. 
The cloak has inner and outer radii $r_0'=0.2$ and $r_1'=0.4$, respectively. 
The disc inside the cloak consists of the same elastic material as the
outer medium but different density. 
We further consider a harmonic unit concentrated force applied either in the
direction $x_1$ or $x_2$ which vibrates with an angular frequency
$\omega=40$. This force is sometimes located outside the
cloak (cf. Fig. \ref{fig02}-\ref{fig04}),
sometimes inside the coating (cf. Fig. \ref{fig05}) or within the central disc (cf. Fig. \ref{fig06}), depending upon whether we are
looking for some neutrality feature, lensing/mirage effect or some
localization.

Before we start looking at the cloaks's features depending upon the
ranges of values of the parameter $\alpha$, we briefly discuss the
implementation of elastic perfectly matched layers (PMLs), in the
framework of transformation elasticity.

\subsection{Implementation of elastic cylindrical PMLs}
\begin{figure}
\centering
\includegraphics[width=6.2cm]{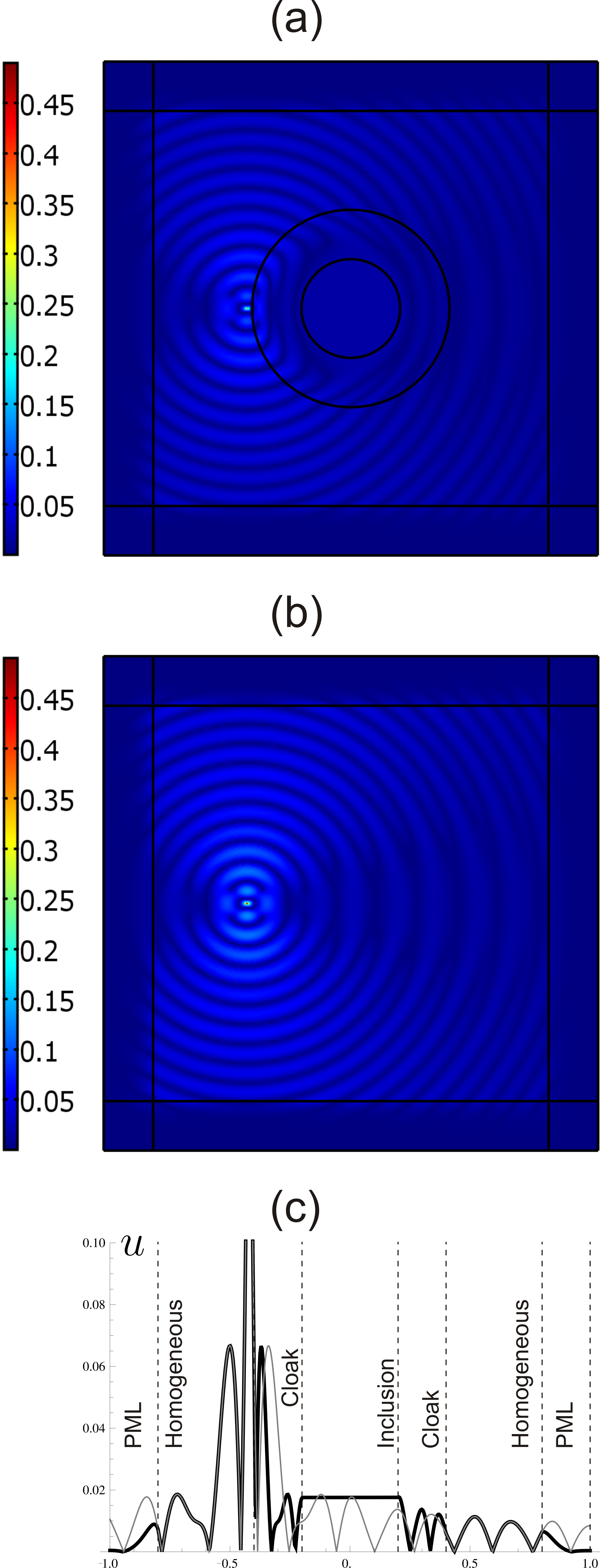}
\vspace{0.2cm} 
\caption{Elastic field generated by an horizontal unit force applied in the external homogenous region; $\alpha=-1+r'_0/r'_1=-0.5$, 
$\omega=40$, source ${\bf x}_0=(-0.42,0)$. 
Magnitude $u=\sqrt{u_1^2+u_2^2}$ of the displacement field in the system with inclusion and cloaking (a) and in an homogenous system (b). Comparison between  the displacement magnitude $u$ computed in Comsol for a cloaked inclusion (black line) and the analytical 
Green's function in an infinite homogeneous linear elastic and isotropic material (gray line), see Eq. (\ref{eqn100}).} 
\label{fig02}
\end{figure}

\begin{figure}
\centering
\includegraphics[width=6.cm]{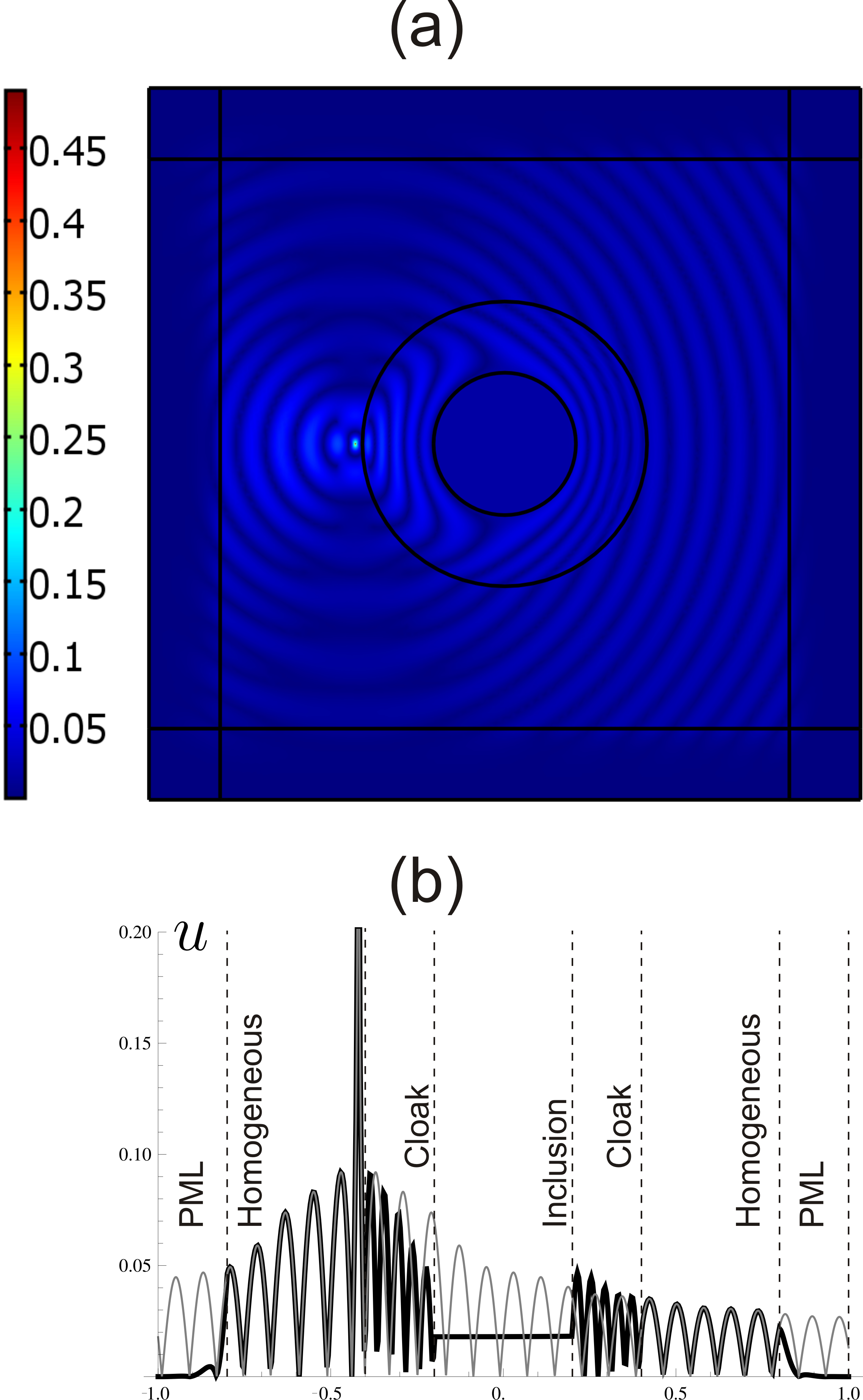}
\vspace{0.2cm} \caption{Elastic field generated by a vertical unit force applied in the external homogeneous region.
$\alpha=-1+r'_0/r'_1=-0.5$, $\omega=40$, source ${\bf x}_0=(-0.42,0)$. 
(a) Magnitude $u$ of the displacement field; (b) Comparison between 
the displacement magnitude $u$ computed in Comsol for a cloaked inclusion (black line) and the analytical 
Green's function in an infinite homogeneous linear elastic and isotropic material (gray line).} 
\label{fig03}
\end{figure}

A perfectly matched layer has been implemented in order
to model the infinite elastic medium surrounding the cloak (cf.
outer ring in Figs. \ref{fig02}-\ref{fig06}); this has been obtained by
application of the geometric transform \cite{zh02},
\begin{equation}
x_i''=(1+a)\hat x_i-ax_i, \qquad i=1,2, 
\end{equation}
for $|x_i|>|\hat{x}_i|$, where $a$ is now a complex number whose imaginary part accounts for
the decay of the elastic waves and $\hat{x}_i=\pm 1$ in Fig. \ref{fig02}-\ref{fig06}. The corresponding
(complex) density $\rho'''$ and elasticity tensor $\cC'''$ are still
given by (\ref{srho}) and (\ref{sc}). The accuracy of the PMLs has
been numerically validated when $a=i-1$, by comparison with the
Green's function in homogeneous elastic space (cf. Eq. \ref{eqn100} and Fig. \ref{fig02}b, c).

We can therefore confidently carry out computations with these PMLs.
We start by the study of an invisibility cloak for in-plane elastic
waves, whereby the point source considered in \cite{brunapl} now
lies in the close vicinity of the cloak (intense near field limit
when the acoustic ray picture breaks down).

\subsection{Neutrality for a point source outside the cloak}
We report in Fig. \ref{fig02} and Fig. \ref{fig03} the computations for a point force applied at a 
distance $r=0.42$ away from the center of the cloak and close to the cloak itself of outer
radius $r'_1=0.4$. 
The force is applied in the horizontal direction in  Fig. \ref{fig02} and in the vertical direction in  Fig. \ref{fig03}.
In both upper panels (a), we clearly see that
both the wave patterns of the magnitude of the displacement field are
smoothly bent around the central region within the cloak (where the
magnitude is uniform).

The comparative analyses between panels (a) and (b) of Fig. \ref{fig02} shows that the wave patterns in the external homogeneous domain 
is not perturbed by the presence of the inclusion and cloaking interface.
This is verified quantitatively in Fig. \ref{fig02} panel (c) and in Fig.  \ref{fig03} panel (b) where the numerically computed wave pattern is compared with the Green's function in homogeneous elastic space
\begin{equation}
\label{eqn100}
{\bf G}({\bf x})\!=\!\frac{i}{4\mu}
\left\{ H_0^{(1)}(k_s r){\bf I}\!-\!\frac{Q}{\omega^2}
\nabla\nabla\left[H_0^{(1)}(k_s r)\!-\!H_0^{(1)}(k_p r)\right] \right\},
\end{equation}
with $H_0^{(1)}$ the Hankel function, $\bf I$ the second order identity tensor, $\nabla$ the gradient operator,
$k_p=\omega/c_p$, $k_s=\omega/c_s$, $Q=(1/c_p^2+1/c_s^2)^{-1}(\lambda+\mu)/(\lambda+2\mu)$,  $c_p=\sqrt{(\lambda+2\mu)/\rho}$, $c_s=\sqrt{\mu/\rho}$.
The plot are reported along the horizontal line $x_2=0$ passing from the point of application of the force.
The absence of forward or backward scattering is demonstrated by the excellent agreement between the two fields in the external homogeneous domain $r>0.4$. Clearly, the profile is much
different when the coating is removed and the inner disc is clamped
or freely vibrating. We also see that the field in the cloaking region has the same amplitude as the one in homogeneous case but shifted following the transformation $r'(r)$; in the inner disk the field is homogeneous.
Finally the effectiveness of the PML domains can also be appreciated.

\begin{figure}
\centering
\includegraphics[width=10cm]{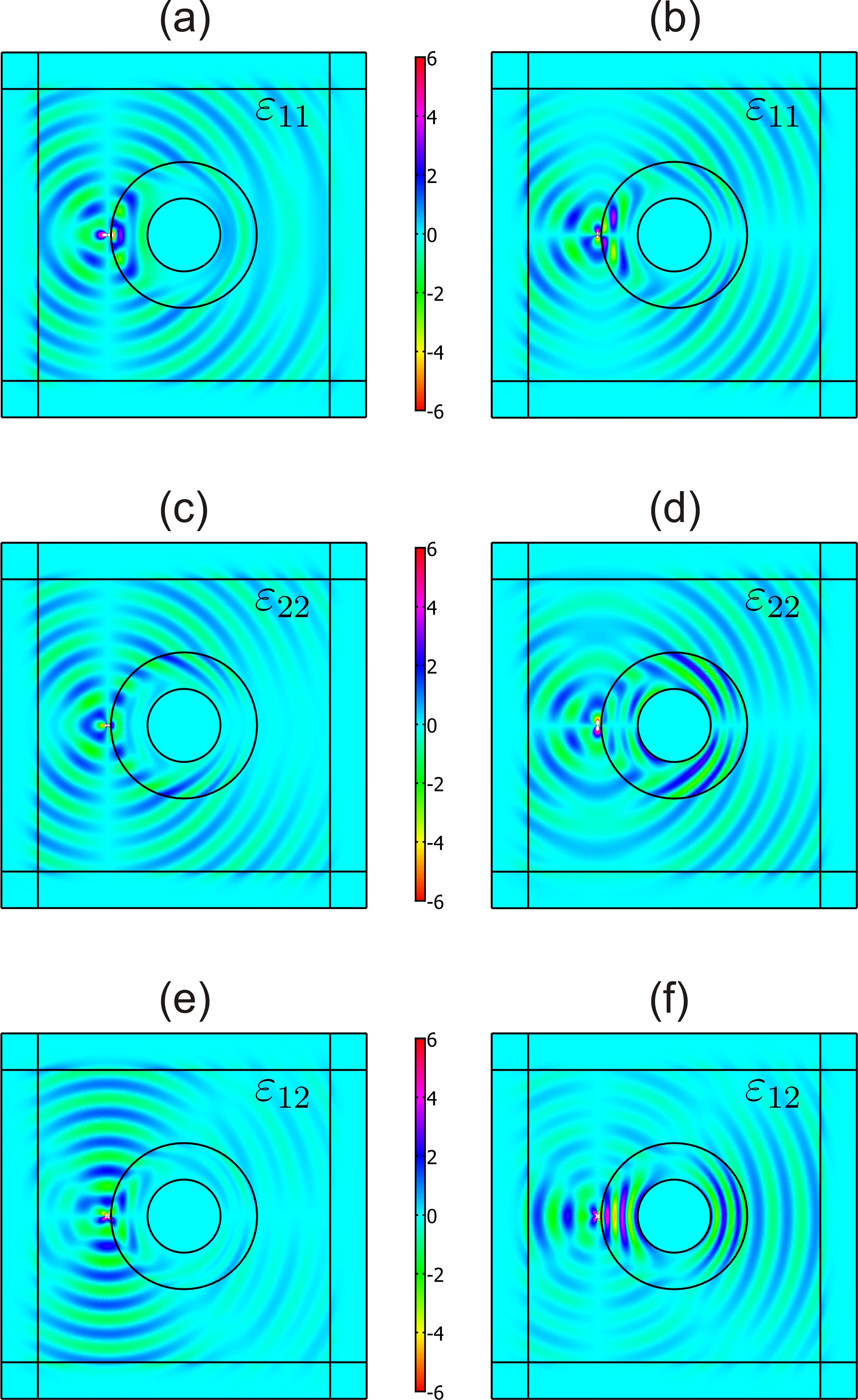}
\vspace{0.2cm} \caption{Elastic deformation fields generated by a unit force applied in the the external homogeneous region. 
$\alpha=-1+r'_0/r'_1=-0.5$, $\omega=40$, source$=(-0.42,0)$ . 
(a), (c), (e) Force applid in the horizontal direction $x_1$. (b), (d), (f) Force applied in the vertical direction $x_2$. 
(a), (b) Component $\varepsilon_{11}=\frac{\partial u_1}{\partial x_1}$; 
(c), (d) Component  $\varepsilon_{22}=\frac{\partial u_2}{\partial x_2}$;
(e), (f) Component  $\varepsilon_{12}=\varepsilon_{21}=\frac{1}{2}(\frac{\partial
u_1}{\partial x_2}+\frac{\partial u_2}{\partial x_1})$.} 
\label{fig04}
\end{figure}

In Fig. \ref{fig04} the deformation fields are also reported for both cases, where the force is applied in horizontal (first column) and vertical (second column) direction.
In the upper (a,b) and central (c,d) panels the skew-symmetric nature of the components
$\varepsilon_{11}$ and $\varepsilon_{22}$ of the deformation tensor reveals the tensor nature of the
problem. The component $\varepsilon_{12}$ leads to a non-intuitive
pattern whereby fully-coupled shear and pressure components create
the optical illusion of interferences. 
Again, the effect of the cloaking is shown and also for deformation fields waves are bent around the cloaking region without backward and forward scattering.

\subsection{Mirage effect for a point source in the coating}
\begin{figure}
\centering
\includegraphics[width=7cm]{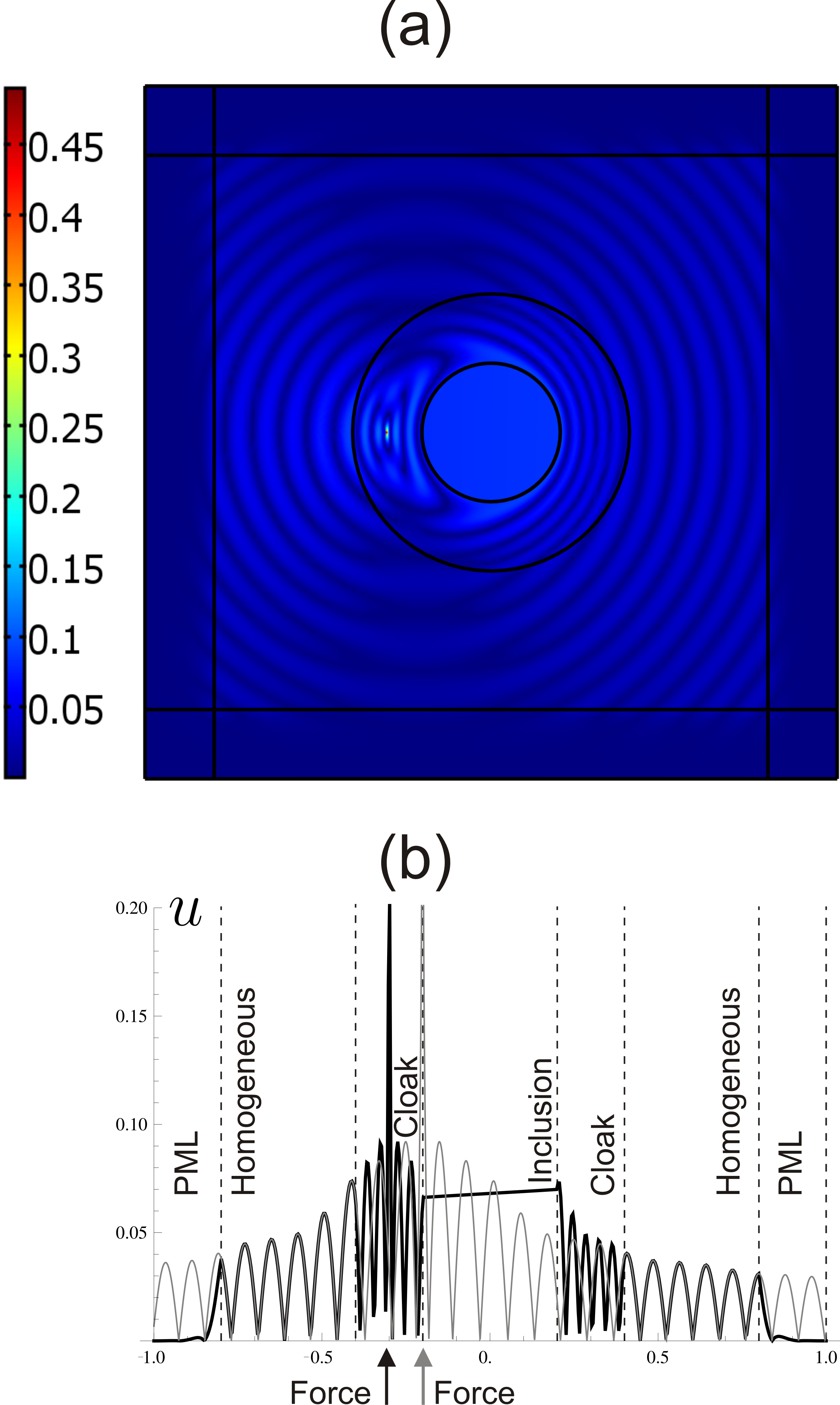}
\vspace{0.2cm} \caption{Elastic field generated by a vertical unit force applied in the cloaking region.
$\alpha=-1+r'_0/r'_1=-0.5$, $\omega=40$, source ${\bf x}_0=(-0.3,0)$. 
(a) Magnitude $u$ of the displacement field; (b) Comparison between 
the displacement magnitude $u$ computed in Comsol for a cloaked inclusion (black line) and the analytical 
Green's function in an infinite homogeneous linear elastic and isotropic material (gray line), corresponding to a force applied in a shifted source point ${\bf x}_0=(-.2,0)$.} 
\label{fig05}
\end{figure}

\begin{figure}
\centering
\includegraphics[width=7cm]{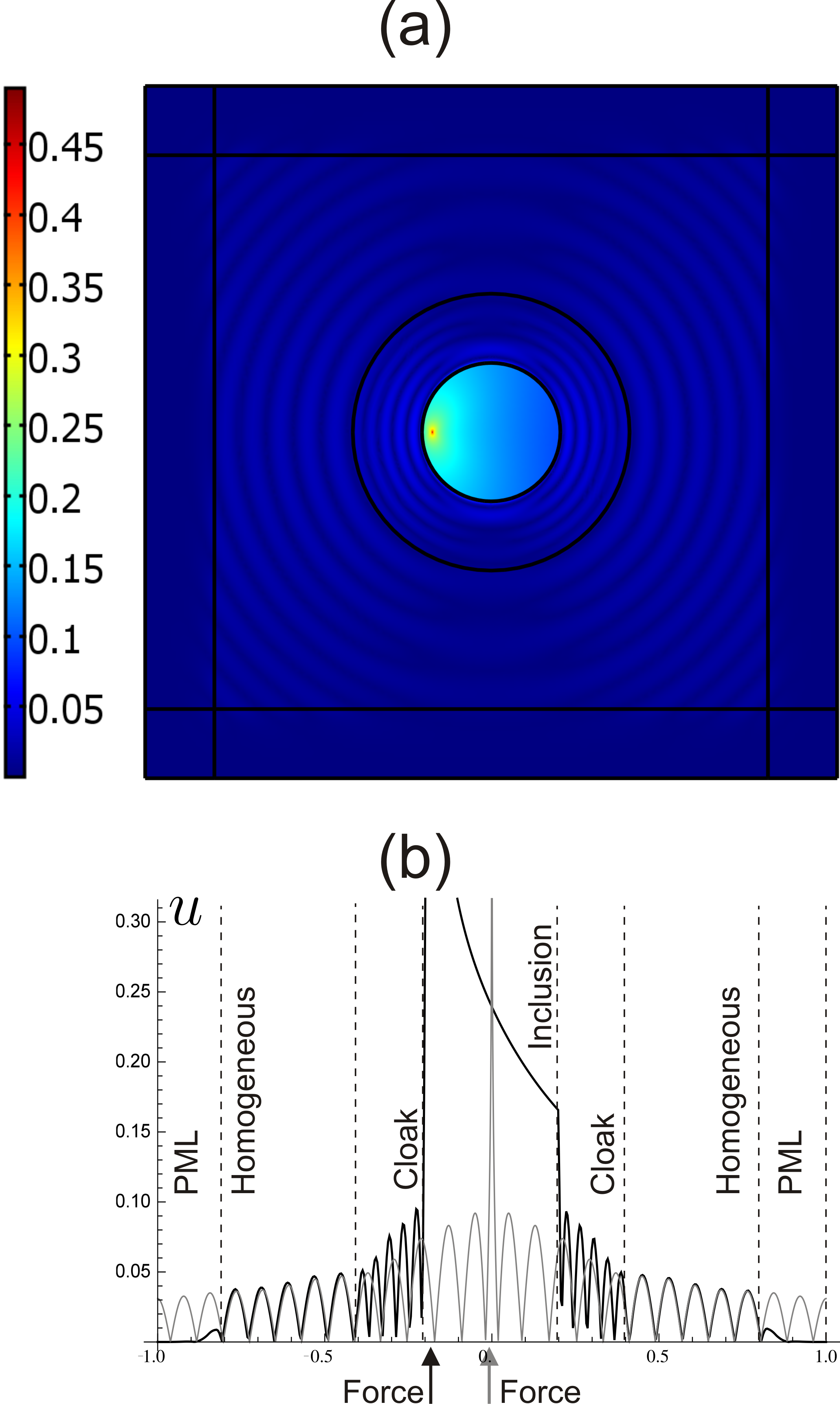}
\vspace{0.2cm} \caption{Elastic field generated by a vertical unit force applied in the internal inclusion.
$\alpha=-1+r'_0/r'_1=-0.5$, $\omega=40$, source ${\bf x}_0=(-0.17,0)$. 
(a) Magnitude $u$ of the displacement field; (b) Comparison between 
the displacement magnitude $u$ computed in Comsol for a cloaked inclusion (black line) and the analytical 
Green's function in an infinite homogeneous linear elastic and isotropic material (gray line), corresponding to a force applied at the origin ${\bf x}_0=(0,0)$.} 
\label{fig06}
\end{figure}

In this section, we look at the case of a point force located inside
the coating. In a way similar to what was observed for an
electromagnetic circular cylindrical cloak \cite{zolla07}, we
observe in Fig. \ref{fig05} a mirage effect: the point force seems to radiate from a location
shifted towards the inner boundary (further away from an observer)
as given by
\begin{equation}
r=\frac{(1+\alpha)r_1-r'}{\alpha} \; , \; \theta=\theta' \;,
\label{invPTransform}
\end{equation}
as also shown in panel (b).

Importantly, the profile of the shifted point source in homogeneous
elastic space is superimposed with that of the point source located
inside the coating, but only outside the cloak. In the invisibility
region i.e. the disc at the center of the cloak, the field is
constant and this suggest that the central region behaves as a
cavity. We study this cavity phenomenon in the next section.

The example in Fig. \ref{fig05} reveals that any object located inside the coating would appear as a
different elastic material with a different shape to an observer.


\subsection{Confinement for a point source in the central region}
We now consider a point force inside the invisibility region. 
Interestingly, a point source located in the invisibility zone always radiates outside the
cloak as if it was located at the origin and this is quite natural
as the central disc is simply the image of the origin via the
geometric transform (\ref{invPTransform}). The fact that the central
disc behaves as a closed cavity is also intuitive, as the elasticity
tensor $\cC'$ is singular on the boundary of the disc.


\subsection{Squeezing the wavelength 
with an elastic concentrator}

\begin{figure}
\centering
\includegraphics[width=7cm]{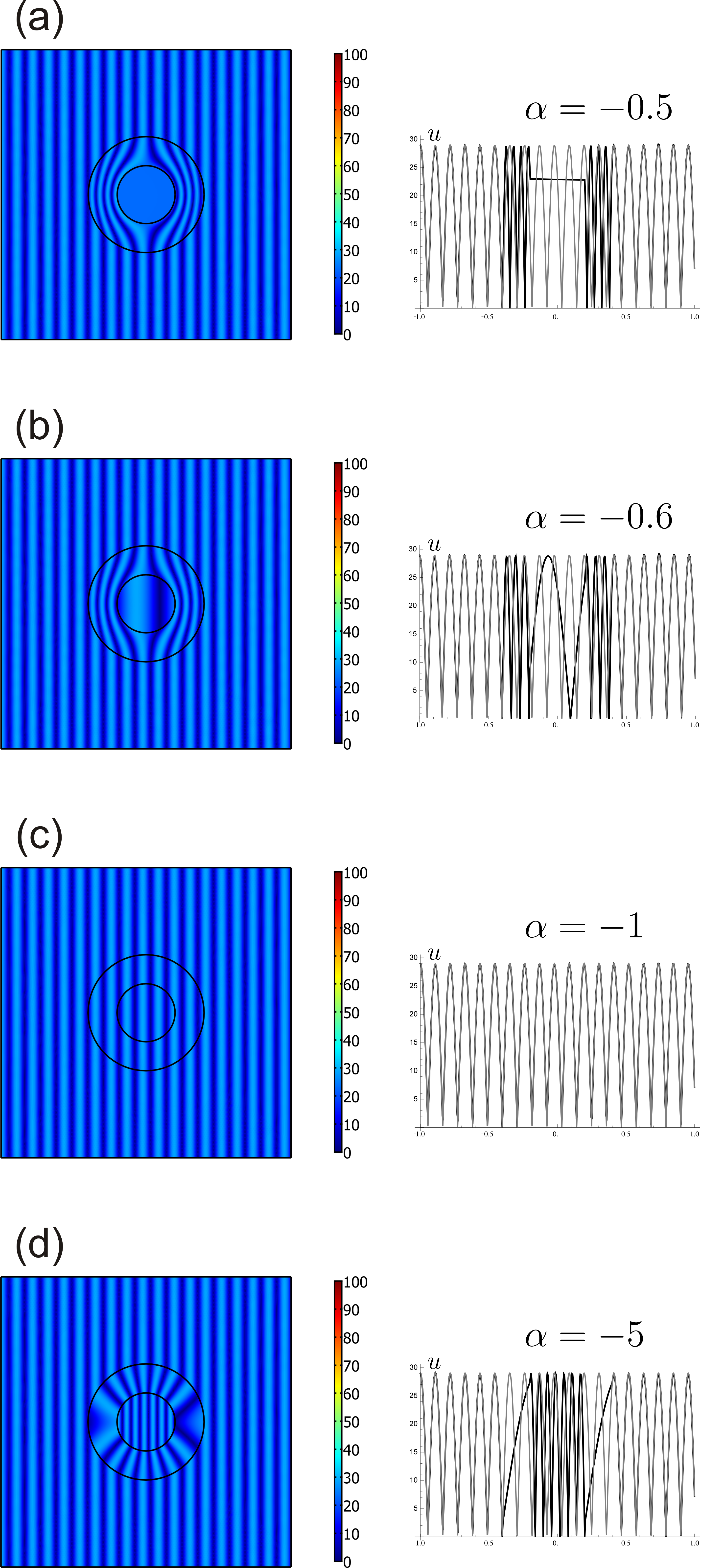}
\vspace{0.2cm} \caption{Elastic field generated by a pressure plane wave ${\bf u}=(A \exp(ik_pt),0)$ with $\omega=60$. Left column: magnitude $u$ of the displacement field. Right column: comparison between  the displacement magnitude $u$ computed in Comsol for a cloaked inclusion (black line) and the pressure plane wave in an infinite homogeneous linear elastic and isotropic material (gray line), results are plotted along an horizontal line passing from the center of the inclusion. (a) $\alpha=-0.5$, (b) $\alpha=-0.6$, (c) $\alpha=-1$, (d) $\alpha=-5$} \label{fig07}
\end{figure}

We report the effects associated to an increase in the magnitude of the parameter $\alpha$ describing the linear transformation (\ref{PTransform}). In Fig. \ref{fig07} the effect of the cylindrical coating on the inclusion is given for a pressure plane wave ${\bf u}=(A \exp(ik_pt),0)$ propagating in the horizontal direction $x_1$. A decrease of $\alpha$ from $\alpha=-1+r'_0/r'_1=-0.5$ introduces wave propagation within the inclusion with progressive shorter wavelengths while the amplitude of the wave remains unchanged. From Fig. \ref{fig07}d it is evident that, when $\alpha<-1$, the interface act as an energy concentrator within the inclusion increasing the energy flux. The energy crossing the inclusion region $r\le r'_0=0.2$ equals the energy crossing the larger region $r\le r_0$, in a homogeneous material. In the interval $-1>\alpha>-\infty$, $r'_0<r_0<r'_1$.

We also note that, when $\alpha\neq -0.5$ the transformation is regular and material parameters remain bounded indicating additional advantages in technological and numerical implementations of the model. Last but not least, the field in the external domain remains unchanged.

\begin{figure}
\centering
\includegraphics[width=7cm]{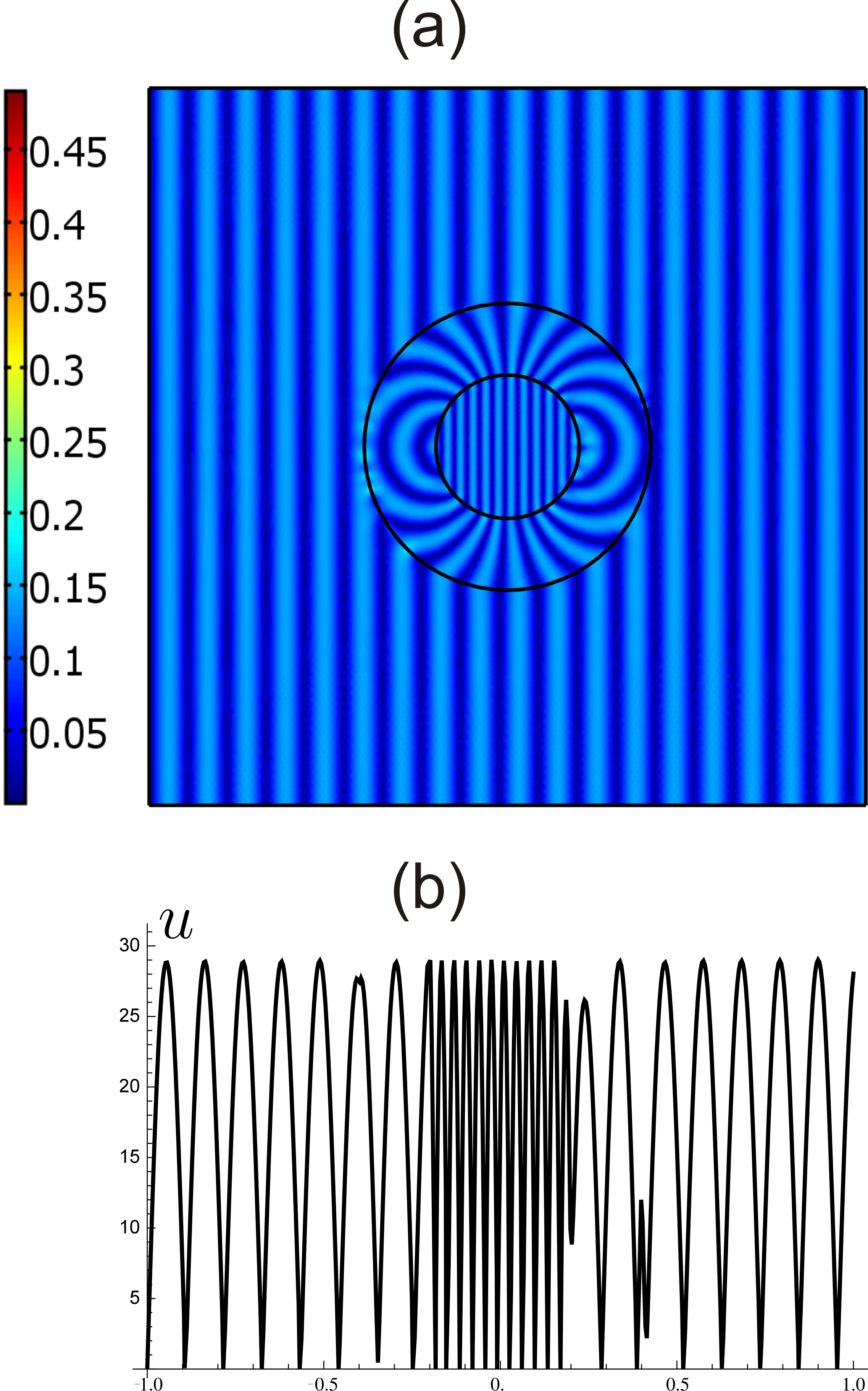}
\vspace{0.2cm} \caption{Elastic field generated by a pressure plane wave ${\bf u}=(A \exp(ik_pt),0)$ with $\omega=60$ and $\alpha=0.94$. (a) Magnitude $u$ of the displacement field. (b) Displacement magnitude $u$ computed in Comsol for a cloaked inclusion plotted along an horizontal line passing from the center of the inclusion.} \label{fig08}
\end{figure}

\subsection{Folding transformation. Superconcentration of an elastic wave with a cylindrical lens}

We finally report in Fig. \ref{fig08} an enhanced energy concentration effect obtained from a folding transformation ($\alpha>0$). In such a case all the energy crossing a circular region larger than the region delimited by the cloaking interface is concentrated into the inner inclusion. In Fig. 8, $\alpha=0.94$ and the radius of the circular region in the homogenous space is $3.06$ times the radius of the inner inclusion. 
Again, such an effect is obtained by an increase in the energy flux leaving unperturbed both the wave amplitude and the fields in the external region.

\section{Conclusion}
We have proposed to use stretched coordinates in order to design an
elastic cloak bending the trajectory of in-plane coupled shear and
pressure waves around an obstacle, concentrating them in its core,
or focussing them. We investigated the transformed equations
of motion for the Mitlon-Briane-Willis transformation gauge ${\bf u}'(r',\theta')={\bf F}^{-t}{\bf u}(r,\theta)$
and the Brun-Guenneau-Movchan gauge ${\bf u}'(r',\theta')={\bf u}(r,\theta)$ (see \cite{NorrisShuvalov2011}). The former leads
to Willis's equations with more extreme anisotropic parameters in the cloak than
for the latter. However, the latter requires a transformed elasticity tensor without
the minor symmetries, which is another hurdle for a metamaterial design.

We have studied various limiting cases for the value of a parameter in the considered
radially symmetric linear geometric transforms. These transforms are applied to the design
of neutral (invisibility) cloaks, elastic concentrators or cylindrical lenses. 

We have numerically explored all the above for the gauge ${\bf u}'(r',\theta')={\bf u}(r,\theta)$
leading to a non-fully symmetric transformed elasticity tensor,
and have notably shown that a source located inside the
anisotropic heterogeneous elastic coating seems to radiate from a
shifted location, and can also lead to anamorphism.

We believe that our space folding based design of elastic cylindrical lenses can lead to an in-plane counterpart of
the external cylindrical cloak for anti-plane shear waves introduced in \cite{guenneau2021time} and applied periodically in \cite{Meirbekova2020}.

We hope our results might open new vistas in cloaking devices for elastodynamic waves. Whereas
their governing equations do not generally retain their form under
geometric transforms, unlike for electromagnetic and acoustic waves, one can choose specific gauges
that can make the transformed equations of motions easier to handle. 

\bibliography{biblio}

\begin{thebibliography}{10}
\expandafter\ifx\csname url\endcsname\relax
  \def\url#1{\texttt{#1}}\fi
\expandafter\ifx\csname urlprefix\endcsname\relax\def\urlprefix{URL }\fi
\expandafter\ifx\csname href\endcsname\relax
  \def\href#1#2{#2} \def\path#1{#1}\fi

\bibitem{farhat09}
M.~Farhat, S.~Guenneau, S.~Enoch, A.~B. Movchan, Cloaking bending waves
  propagating in thin elastic plates, Physical Review B 79~(3) (2009) 033102.

\bibitem{farhat09PRL}
M.~Farhat, S.~Guenneau, S.~Enoch, Ultrabroadband elastic cloaking in thin
  plates, Physical review letters 103~(2) (2009) 024301.

\bibitem{stenger12}
N.~Stenger, M.~Wilhelm, M.~Wegener, Experiments on elastic cloaking in thin
  plates, Physical Review Letters 108~(1) (2012) 014301.

\bibitem{misseroni16}
D.~Misseroni, D.~J. Colquitt, A.~B. Movchan, N.~V. Movchan, I.~S. Jones,
  Cymatics for the cloaking of flexural vibrations in a structured plate,
  Scientific reports 6~(1) (2016) 23929.

\bibitem{colquitt14}
D.~J. Colquitt, M.~Brun, M.~Gei, A.~B. Movchan, N.~V. Movchan, I.~S. Jones,
  Transformation elastodynamics and cloaking for flexural waves, Journal of the
  Mechanics and Physics of Solids 72~(1) (2014) 131--143.

\bibitem{pomot2019form}
L.~Pomot, S.~Bourgeois, C.~Payan, M.~Remillieux, S.~Guenneau, On form
  invariance of the kirchhoff-love plate equation, arXiv preprint
  arXiv:1901.00067 (2019).

\bibitem{Brun14b}
M.~Brun, D.~J. Colquitt, I.~S. Jones, A.~B. Movchan, N.~N. Movchan,
  Transformation cloaking and radial approximations for flexural waves in
  elastic plates, New Journal of Physics 16 (2014) 093020.

\bibitem{morvaridi18}
M.~Morvaridi, M.~Brun, Perfectly matched layers for flexural waves in
  kirchhof-love plates, International Journal of Solids and Structures 134
  (2018) 293--303.

\bibitem{Golgoon21}
A.~Golgoon, A.~Yavari, Transformation cloaking in elastic plates, Journal of
  Nonlinear Science 31~(17) (2021) 1--76.

\bibitem{rayoptics}
D.~Schurig, J.~Pendry, D.~R. Smith, Calculation of material properties and ray
  tracing in transformation media, Optics express 14~(21) (2006) 9794--9804.

\bibitem{leonhardt06}
U.~Leonhardt, Optical conformal mapping, science 312~(5781) (2006) 1777--1780.

\bibitem{pendry2006controlling}
J.~B. Pendry, D.~Schurig, D.~R. Smith, Controlling electromagnetic fields,
  science 312~(5781) (2006) 1780--1782.

\bibitem{greenleaf03}
A.~Greenleaf, M.~Lassas, G.~Uhlmann, Anisotropic conductivities that cannot be
  detected by eit, Physiological measurement 24~(2) (2003) 413.

\bibitem{greenleafprl07}
A.~Greenleaf, Y.~Kurylev, M.~Lassas, G.~Uhlmann, Electromagnetic wormholes and
  virtual magnetic monopoles from metamaterials, Physical Review Letters
  99~(18) (2007) 183901.

\bibitem{kadic2014invisible}
M.~Kadic, G.~Dupont, S.~Enoch, S.~Guenneau, Invisible waveguides on metal
  plates for plasmonic analogs of electromagnetic wormholes, Physical Review A
  90~(4) (2014) 043812.

\bibitem{philbin08}
T.~G. Philbin, C.~Kuklewicz, S.~Robertson, S.~Hill, F.~K{\"o}nig, U.~Leonhardt,
  Fiber-optical analog of the event horizon, Science 319~(5868) (2008)
  1367--1370.

\bibitem{hawking}
S.~W. Hawking, Black hole explosions?, Nature 248~(5443) (1974) 30--31.

\bibitem{pendry1999magnetism}
J.~B. Pendry, A.~J. Holden, D.~J. Robbins, W.~Stewart, Magnetism from
  conductors and enhanced nonlinear phenomena, IEEE transactions on microwave
  theory and techniques 47~(11) (1999) 2075--2084.

\bibitem{schurig06}
D.~Schurig, J.~J. Mock, B.~J. Justice, S.~A. Cummer, J.~B. Pendry, A.~F. Starr,
  D.~R. Smith, Metamaterial electromagnetic cloak at microwave frequencies,
  Science 314~(5801) (2006) 977--980.

\bibitem{zhangprl08}
S.~Zhang, D.~A. Genov, C.~Sun, X.~Zhang, Cloaking of matter waves, Physical
  Review Letters 100~(12) (2008) 123002.

\bibitem{greenleafprl08}
A.~Greenleaf, Y.~Kurylev, M.~Lassas, G.~Uhlmann, Approximate quantum cloaking
  and almost-trapped states, Physical review letters 101~(22) (2008) 220404.

\bibitem{cummer06b}
S.~A. Cummer, D.~Schurig, One path to acoustic cloaking, New Journal of Physics
  9~(3) (2007) 45.

\bibitem{chen07}
H.~Chen, C.~Chan, Acoustic cloaking in three dimensions using acoustic
  metamaterials, Applied physics letters 91~(18) (2007) 183518.

\bibitem{cummer08}
S.~A. Cummer, B.-I. Popa, D.~Schurig, D.~R. Smith, J.~Pendry, M.~Rahm,
  A.~Starr, Scattering theory derivation of a 3d acoustic cloaking shell,
  Physical review letters 100~(2) (2008) 024301.

\bibitem{sanchez}
D.~Torrent, J.~S{\'a}nchez-Dehesa, Acoustic metamaterials for new
  two-dimensional sonic devices, New journal of physics 9~(9) (2007) 323.

\bibitem{farhat08}
M.~Farhat, S.~Guenneau, S.~Enoch, A.~B. Movchan, Cloaking bending waves
  propagating in thin elastic plates, Physical Review B 79~(3) (2009) 033102.

\bibitem{milton06b}
G.~W. Milton, M.~Briane, J.~R. Willis, On cloaking for elasticity and physical
  equations with a transformation invariant form, New Journal of Physics 8~(10)
  (2006) 248.

\bibitem{bigoni98}
D.~Bigoni, S.~Serkov, M.~Valentini, A.~Movchan, Asymptotic models of dilute
  composites with imperfectly bonded inclusions, International journal of
  solids and structures 35~(24) (1998) 3239--3258.

\bibitem{willis1981variational}
J.~R. Willis, Variational principles for dynamic problems for inhomogeneous
  elastic media, Wave Motion 3~(1) (1981) 1--11.

\bibitem{willis1985nonlocal}
J.~R. Willis, The nonlocal influence of density variations in a composite,
  International Journal of Solids and Structures 21~(7) (1985) 805--817.

\bibitem{brunapl}
M.~Brun, S.~Guenneau, A.~B. Movchan, Achieving control of in-plane elastic
  waves, Applied physics letters 94~(6) (2009) 061903.

\bibitem{nassar2018degenerate}
H.~Nassar, Y.~Chen, G.~Huang, A degenerate polar lattice for cloaking in full
  two-dimensional elastodynamics and statics, Proceedings of the Royal Society
  A 474~(2219) (2018) 20180523.

\bibitem{Garau2019}
M.~Garau, M.~J. Nieves, G.~Carta, M.~Brun, Transient response of a gyro-elastic
  structured medium: Unidirectional waveforms and cloaking, International
  Journal of Engineering Science 143 (2019) 115--141.

\bibitem{achaoui2020cloaking}
Y.~Achaoui, A.~Diatta, M.~Kadic, S.~Guenneau, Cloaking in-plane elastic waves
  with swiss rolls, Materials 13~(2) (2020) 449.

\bibitem{diatta2014controlling}
A.~Diatta, S.~Guenneau, Controlling solid elastic waves with spherical cloaks,
  Applied Physics Letters 105~(2) (2014) 021901.

\bibitem{parnell2012nonlinear}
W.~J. Parnell, Nonlinear pre-stress for cloaking from antiplane elastic waves,
  Proceedings of the Royal Society A: Mathematical, Physical and Engineering
  Sciences 468~(2138) (2012) 563--580.

\bibitem{norris2012hyperelastic}
A.~N. Norris, W.~J. Parnell, Hyperelastic cloaking theory: transformation
  elasticity with pre-stressed solids, Proceedings of the Royal Society A:
  Mathematical, Physical and Engineering Sciences 468~(2146) (2012) 2881--2903.

\bibitem{parnell2012employing}
W.~J. Parnell, A.~N. Norris, T.~Shearer, Employing pre-stress to generate
  finite cloaks for antiplane elastic waves, Applied Physics Letters 100~(17)
  (2012) 171907.

\bibitem{rahm}
M.~Rahm, D.~Schurig, D.~A. Roberts, S.~A. Cummer, D.~R. Smith, J.~B. Pendry,
  Design of electromagnetic cloaks and concentrators using form-invariant
  coordinate transformations of maxwell's equations, Photonics and
  Nanostructures-fundamentals and Applications 6~(1) (2008) 87--95.

\bibitem{NorrisShuvalov2011}
A.~N. Norris, A.~L. Shuvalov, Elastic cloaking theory, Wave Motion 48~(6)
  (2011) 525--538.

\bibitem{Ogden1997}
R.~W. Ogden, Non-linear elastic deformations, Courier Corporation, 1997.

\bibitem{Kohn2008}
R.~V. Kohn, H.~Shen, M.~S. Vogelius, M.~I. Weinstein, Cloaking via change of
  variables in electric impedance tomography, Inverse Problems 24~(1) (2008)
  015016.

\bibitem{Colquittetal2013}
D.~Colquitt, I.~Jones, N.~Movchan, A.~Movchan, M.~Brun, R.~McPhedran, Making
  waves round a structured cloak: lattices, negative refraction and fringes,
  Proceedings of the Royal Society A: Mathematical, Physical and Engineering
  Sciences 469~(2157) (2013) 20130218.

\bibitem{Jonesetal2015}
I.~Jones, M.~Brun, N.~Movchan, A.~Movchan, Singular perturbations and cloaking
  illusions for elastic waves in membranes and kirchhoff plates, International
  journal of Solids and Structures 69 (2015) 498--506.

\bibitem{Miltonetal2008}
G.~W. Milton, N.-A.~P. Nicorovici, R.~C. McPhedran, K.~Cherednichenko,
  Z.~Jacob, Solutions in folded geometries, and associated cloaking due to
  anomalous resonance, New Journal of Physics 10~(11) (2008) 115021.

\bibitem{zh02}
Y.~Zheng, X.~Huang, Anisotropic perfectly matched layers for elastic waves in
  cartesian and curvilinear coordinates, Tech. rep., Massachusetts Institute of
  Technology. Earth Resources Laboratory (2002).

\bibitem{zolla07}
F.~Zolla, S.~Guenneau, A.~Nicolet, J.~Pendry, Electromagnetic analysis of
  cylindrical invisibility cloaks and the mirage effect, Optics Letters 32~(9)
  (2007) 1069--1071.

\bibitem{guenneau2021time}
S.~Guenneau, B.~Lombard, C.~Bellis, Time-domain investigation of an external
  cloak for antiplane elastic waves, Applied Physics Letters 118~(19) (2021)
  191102.

\bibitem{Meirbekova2020}
B.~Meirbekova, M.~Brun, Control of elastic shear waves by periodic geometric
  transformation: cloaking, high reflectivity and anomalous resonances, Journal
  of the Mechanics and Physics of Solids 137 (2020) 103816.

\end{thebibliography}

\end{document}